# Landau-Levich Scaling for Optimization of Quantum Dot Layer Morphology and Thickness for Enhanced Quantum-Dot Light-Emitting Diode Performance

Yiman Xu[1], Grant J. Dixon[2], Qing Xie[3], James F. Gilchrist[1], Brandi M. Cossairt[2], David S. Ginger[2], Elsa Reichmanis[1]*

1. Department of Chemical and Biomolecular Engineering, Lehigh University, 124 E. Morton Street, Bethlehem, Pennsylvania 18015, United States
2. Department of Chemistry, University of Washington, Seattle, Washington 98195, United States
3. Department of Chemistry, Lehigh University, 6 E Packer Avenue, Bethlehem, Pennsylvania 18015, United States

**Abstract**

Quantum dot (QD) light-emitting diodes (QLEDs) are promising candidates for next-generation displays because of their high efficiency, brightness, broad color gamut, and solution-processability. Large-scale solution-processing of electroluminescent QLEDs poses significant challenges, particularly concerning the precise control of the active layer's thickness and uniformity. These obstacles directly impact charge transport, leading to current leakage and reduced overall efficiency. Blade-coating is a prevalent and scalable solution processing technique known for its speed and minimal waste. Additionally, it allows for continuous "roll-to-roll" processing, making it highly adaptable in various applications. In this study, we demonstrate the precise control of blade speed in the Landau-Levich regime to create a uniform QD emission layer, using a commercial CdSe/ZnS QD as a representative example. Blade coating provides QDs with time to reorganize as solvent evaporation occurs in the latter stage of the Landau-Levich regime. QDs assemble into different morphologies on glass and the underlying layers of the QLED device due to variations in interaction energy. The QD film thickness can be modified from monolayer to multilayer by adjusting blade speed, which can be predicted by fitting the Landau-Levich-Derjaguin theory. The optimal speed at 7 mm/s results in a QD film with a surface coverage of around 163% and low roughness (1.57 nm mean square height). The QLED external quantum efficiency (EQE) of approximately 1.5% was achieved using commercially available CdSe/ZnS QDs with low photoluminescence quantum yield (PLQY), and an EQE of around 7% has been obtained using lab-made InP/ZnSe/ZnS QDs having a solution PLQY of 74%. All-blade-coated CdSe-QLEDs are further demonstrated by adopting the optimized speed for the QD layer. This method demonstrates significant potential for developing low-cost, reproducible, and scalable QLED technologies with uniform emission characteristics and low-waste production.

**Introduction**

Colloidal quantum dots (QDs) are nano-sized semiconductor particles that exhibit quantum confinement and are stabilized in solvents through ligands. Their distinctive size-dependent optical properties have motivated extensive research in the development of optoelectronic devices.[1–6] Among these, quantum dot light-emitting diodes (QLEDs) are attractive candidates for next-generation displays due to their high efficiency, brightness, wide color gamut, and solution processability.[7–9]

The past decades have seen spin-coated QLEDs with remarkably improved brightness, efficiency, and stability.[10–13] Spin-coating contributes to uniform nanoscale thick films that benefit the electroluminance performance of QLEDs since the film thickness and quality affect the carrier transport efficiency and brightness. However, spin-coating results in significant material waste and is restricted to rigid substrates of limited size, while its batch-to-batch processing approach hampers production efficiency.[14] Therefore, industrial-scale large-area processing techniques for QLEDs have attracted research interest.[15–19]

Blade-coating is a widely used scalable technique for creating thin films. The blade is moved across the substrate, evenly spreading the solution to form a uniform thin film. This technique is low-waste and amenable to continuous "roll-to-roll" processing.[20–22] In general, coating speed and angle, temperature, ink solvents, and concentrations all influence the final thin film quality.[23] Malaquin et al.[24] investigated substrate temperature, velocity, wetting properties, and pattern geometry by assembling gold and polystyrene colloidal nanoparticles. They found that setting the temperature of the colloidal suspension above or below the dew point can control the assembly and disassembly. Using phospholipid molecules, Maël et al.[25] found that coating speed is critical in controlling thin film thickness. At low speeds, phospholipid molecules accumulate near the contact line, forming a dry film behind the meniscus (evaporation regime). At high speed, viscous forces become predominant and pull out a liquid film that will dry afterward. This is known as the Landau−Levich regime, where the liquid film thickness scales with the coating velocity to two-thirds power.[26] Preparing uniform-emitting, high-performance QLEDs by blade coating is still challenging due to film thickness control and uniformity issues. Efforts to achieve uniformity in the blade-coated hole injection layer (HIL) and electron transport layer (ETL) have involved using mixed solvents, adding another solvent to the precursor, and incorporating surfactants.[27–29]

Similarly, the binary-solvent strategy is adopted to improve the uniformity of the bladed-coated perovskite QD film. Zeng et al.[30] adjusted the QD solution concentration to investigate the relationship between blade-coated film morphology and QLED device performance. However, the formation of QLEDs through continuous processing of QD films would be enhanced by applying existing fluid mechanics scaling principles.

In this study, we present a strategy for large-area production of QLEDs by blade-coating using a commercial CdSe/ZnS QD with octadecylamine ligands as a representative example. A systematic investigation of QD deposition at varying speeds was conducted to precisely control thin film morphology and thickness in the Landau-Levich regime. This assembly is affected by the interaction energy between QD-QD ($W_{QD-QD}$) and QD-substrate ($W_{QD-subs}$), resulting in different QD film morphologies on the glass and the underlying layers of the QLED device. Blade-coating allows for slower solvent evaporation, giving QDs more time to reorganize. The QD layer thickness at lower blade speeds fits well with Landau-Levich-Derjaguin theory, which is further used to predict the thickness at higher blade speeds. QLED performance is closely related to the QD assembly. The QLED devices with one and a half layers of blade-coated QD film, corresponding to a base layer of near maximally packed QDs and a second layer of half of the maximum packing, and low roughness demonstrated the highest measured device efficiencies. However, efficiencies decreased as the film thickness increased. The QLED devices with quantum dots blade-coated at 7 mm/s achieved the highest external quantum efficiency (EQE) of 1.5% and luminance of 2397 cd/m$^2$. Furthermore, all-blade-coated QLEDs with uniform emission and low waste have been demonstrated. This technique is further adapted to lab-made InP/ZnSe/ZnS QDs, resulting in a blade-coated QLED device with an EQE of 6.6% and a luminance of about 10,000 cd/m$^2$. The significance of this work lies in achieving precise control over layer thickness at the nanoscale during blade-coating. Additionally, we demonstrate the viability of blade coating as a scalable method for QLED fabrication, paving the way for future optimization with high-quality QDs featuring different ligand chemistry. Further, the approach offers a path toward low-waste, reproducible, and scalable QLED technologies.

**Results and Discussion**

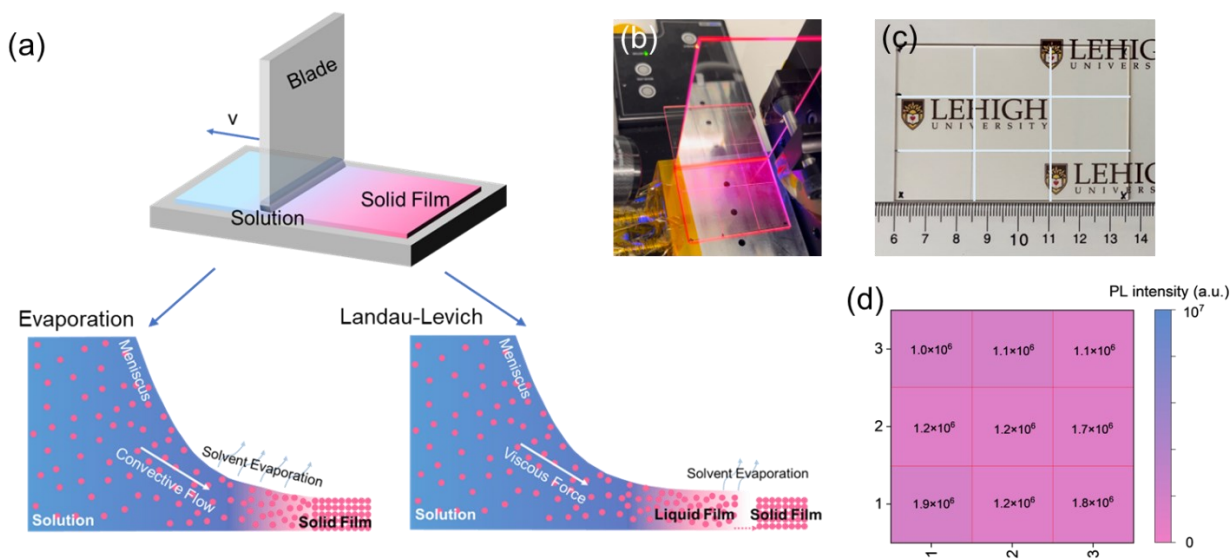

**Figure 1**. (a) A schematic illustrating the blade-coating process, and the fluid dynamics involved in the evaporation and Landau-Levich regimes. (b) A digital image of homemade blade coater processing a large-area QD film under UV light (395 nm). (c) Image of the large-area QD film and (d) its photoluminescence (PL) mapping.

**Figure 1a** schematically shows the QD blade coating process and two principal coating regimes. A blade moves across the substrate at a given speed and drags the QD solution along the substrate. The meniscus is formed between the blade and the substrate. The QD assembly principle can be divided into two regimes - evaporation and Landau-Levich - based on the relative influence of viscous stresses and evaporation driving deposition. In the evaporation regime, low coating speed results in faster evaporation relative to the coating process. QD solution is concentrated near the contact line due to self-filtration through the particle layer, which results from solvent evaporation from the deposited layer. In the Landau-Levich regime, the higher coating speed outpaces the evaporative flux at the interface. The QD solution is dragged out as a liquid film by viscous forces and then transformed into a thinner solid film as the solvent evaporates from the almost stagnant liquid layer downstream. The Landau-Levich regime was chosen because the coating speed is much faster and is suitable for efficient scale-up film printing. We fabricated large-area CdSe/ZnS QD films on glass to demonstrate the robustness of blade coating (**Figure 1b-c**). The QD film was cut into nine pieces for PL measurements, revealing uniform emission across the samples (**Figure 1d**). This example demonstrates the potential for large-area production of uniform QD thin films, an important step toward achieving high-performance QLEDs.

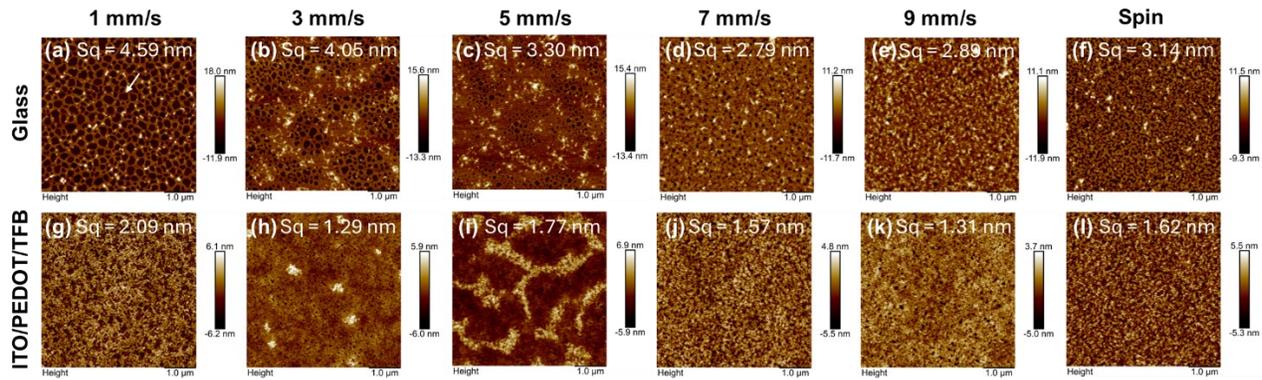

**Figure 2**. The morphology of QD films blade-coated at different speeds (a-f) on glass and (g-j) on ITO/ poly(3,4-ethylenedioxythiophene):poly(styrenesulfonate) (PEDOT)/ poly[(9,9-dioctylfluorenyl-2,7-diyl)-alt(4,4′-(N-(4-butylphenyl)))] (TFB) substrates with different roughness Sq (squared mean height).

Atomic force microscopy (AFM) was used to examine the morphology of CdSe/ZnS QD films that were blade-coated at various speeds, as uniform QD layers are crucial for consistent QLED emission. Here, different substrates were considered given that glass is a typical substrate used to measure film morphology,[31,32] and ITO/PEDOT/TFB is a popular lower layer for bottom emission QLED devices.[33–35] The as-prepared ITO/PEDOT/TFB is uniform with a roughness of 0.561 nm

(**Figure S1a**). The contact angles of the QD solution are less than ten degrees on both substrates, indicating good wettability (**Figure S2**). This wetting is beneficial for achieving a uniform QD layer. In the Landau-Levich regime, the blade speed increases, and the deposited QD film is thicker, which is the same trend for both substrates (**Figure 2a-e, g-k**). However, the QD films show different morphologies on glass *vs*. ITO/PEDOT/TFB because the assembly of the QD film is determined by QD-substrate ($W_{QD-subs}$) and the QD-QD ($W_{QD-QD}$) interaction energy. When $W_{QD-subs} > W_{QD-QD}$, the QDs have a higher affinity for each other than the substrate. Further, QD-QD interactions are influenced by the fluid motion, as highlighted by the images of QDs coated on glass (**Figure 2a-f**). Notably, the honeycomb-like pattern of the QD sub-monolayer formed on glass (pointed out by a white arrow) is due to dewetting of the thin layer of solution in the final stage of drying.[36] As the liquid film on the substrate thins, capillary attraction between the QDs leads to aggregation in the dried films, resulting in QD films having high roughness on glass. Alternatively, with $W_{QD-subs} < W_{QD-QD}$, the QDs have a higher affinity to the substrate. Thus, the QDs are more likely to stick to the substrate surface, which is the case for the ITO/PEDOT/TFB substrate (**Figure 2g-l**). In this case, the QDs are more likely to interact with the lower substrate and form a uniform monolayer. Subsequently, a second layer forms over the first, as exemplified by the pattern shown in **Figure 2i**. This pattern may be a result of Rayleigh–Bénard convection,[37,38] which arises from a temperature gradient created by the cooling effect of evaporation. As the speed continues to increase, more QDs are deposited from a thicker cast film, eventually forming a continuous second layer. The above blade-coated films were further compared with the spin-coated analogs. In the spin coating process, the QD solution is dropped onto a rotating substrate. The rotation generates centrifugal and viscous shear forces, causing the solution to spread and thin across the substrate, flinging excess solution rapidly from the substrate. The initial thickness correlates with the height of the viscous boundary layer that remains. Spinning enhances the mass transfer coefficient, and as the solvent evaporates, a uniform QD thin film is formed on the substrate. The roughness of the blade-coated and spin-coated films are comparable on either substrate. The roughness of the blade-coated and spin-coated films are comparable on either substrate (**Figure S3**). Interaction energy also influences spin coating, where higher film roughness on glass due to QD aggregation aligns with the results observed in blade coating.

The thickness and continuity of the QD emission layer strongly influence QLED performance, as it determines whether electrons and holes undergo radiative recombination to produce light or pass through the device, leading to current leakage. We further investigated the thickness of the QD films on the actual device substrate. The Landau-Levich-Derjaguin theory is widely applied to predict the thickness of the thin film coating process in the Landau-Levich regime.[26,39] This theory predicts that the thickness of the liquid layer dragged by the substrate is a function of the coating velocity when viscous forces dominate, as represented by equation 1

$$h_0 = 0.938 \frac{(\eta u)^{\frac{2}{3}}}{(\rho g)^{\frac{1}{2}} \sigma^{\frac{1}{6}}} \qquad (1)$$

where $h_0$ is the thickness of the liquid layer, $\eta$ is fluid viscosity, $u$ is coating velocity, $\rho$ is the fluid density, $g$ is gravitational acceleration, $\sigma$ is the surface tension of the fluid.

We used equation 1 to describe the coverage of QD films on the ITO/PEDOT/TFB substrate. For a fixed concentration, a simple mass balance demonstrates the height of the liquid layer is proportional to the thickness of the dried film (**Figure 3a**). Thus, the surface coverage is a function of the coating velocity for a given solvent with invariant density and viscosity, as described by equation 2,

$$c = Au^{\frac{2}{3}} \tag{2}$$

where $c$ is the percentage of surface coverage and $A$ is a constant.

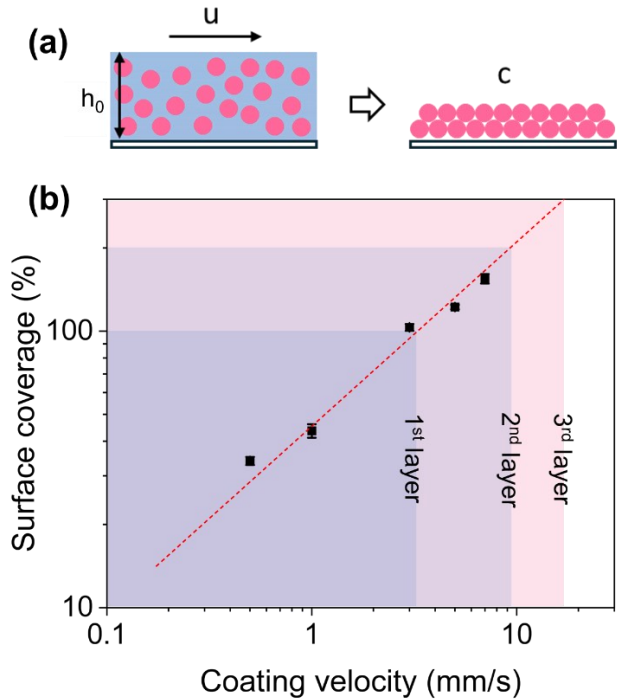

**Figure 3**. (a) The schematic of the liquid film drying in the Landau-Levich regime. (b) The plot of the surface coverage of the QD film on the ITO/PEDOT/TFB substrate as a function of the coating velocity. The red dashed line with a slope of 2/3 is predicted by Landau-Levich scaling. The colored squares illustrate the predicted layer growth at various coating speeds.

With a known size of the CdSe/ZnSe QD of ~8 nm, QD layers coated at 0.5 and 1 mm/s show a clear sub-monolayer morphology (**Figure 2g**, **Figure S1b**), while the films coated at 3 and 5 mm/s have a distinctly visible second layer (**Figure 2h-i**). Therefore, based on AFM imaging, we examined the morphology of QD films processed at low coating speeds to validate alignment with the above theory. By fitting the surface coverage to the coating velocity, the exponent factor is 0.60 ± 0.04, close to 2/3, indicating good agreement with the Landau-Levich-Derjaguin theory. Thus, we used 2/3 as the exponent factor and found $A$ = 45.16 (**Figure 3b**). By fitting in the equation, we can estimate the coverage of other films coated at different velocities (**Table S1**). According to the estimation, films coated at 3-4 mm/s should have monolayer coverage, 9-10 mm/s should afford bilayers, and 17-18 mm/s should lead to trilayer coverage. Therefore, the thickness of the QD layer can be adjusted by simply changing the blade speed.

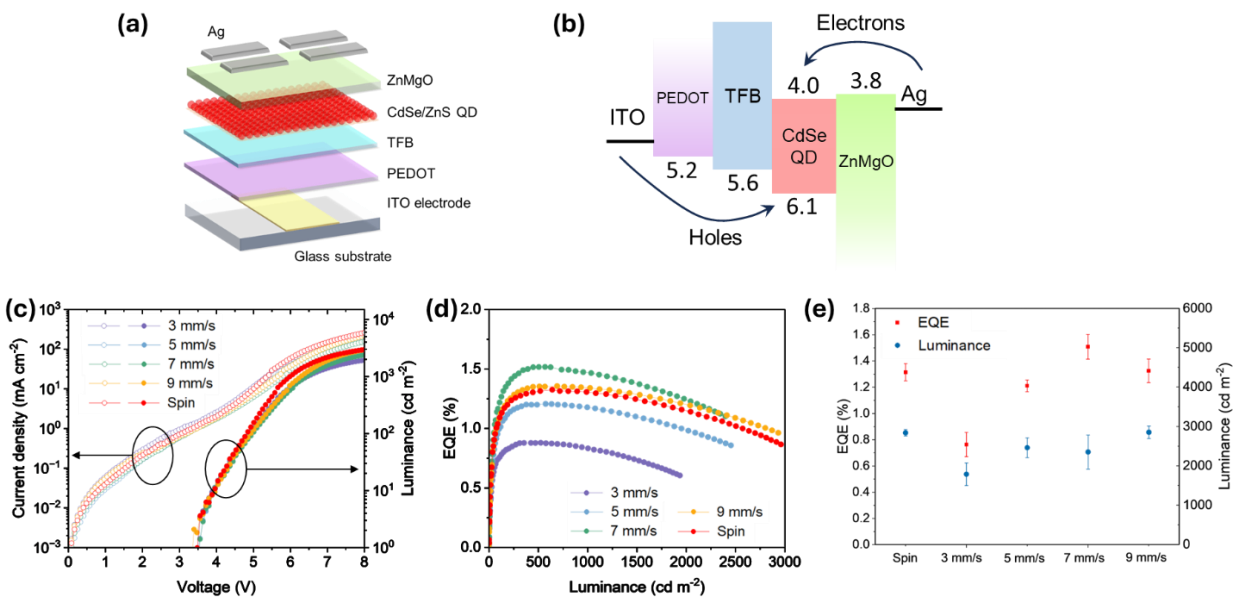

**Figure 4**. (a) The schematic of the CdSe-QLED structure and (b) the energy level diagram for active layers. (c) Current density (J) and luminance (L) versus driving voltage (V) curve. (d) EQE versus L curve. (e) Dependence of CdSe-QLED maximum EQE and L on the processing methods and speeds.

We fabricated CdSe-QLEDs with a structure of ITO/PEDOT/TFB/QD/ZnMgO/Ag (**Figure 3a**). The energy landscape is shown in **Figure 3b**. This device structure is optimal since good electron and hole balance was observed in the spin-coated devices, as evidenced by J-V curves of electron- and hole-only devices (**Figure S4**). Notably, the electroluminescence (EL) spectra for CdSe-QLED devices have two small peaks, which point to the presence of a bimodal distribution in the size of the commercial QD samples (**Figure S5**). The new emission band is prominent at higher voltages, which may come from the 1P transition.[40] To investigate the impact of blade coating on device performance, we compared QLEDs fabricated with QD layers blade-coated at different speeds with fully spin-coated devices (**Figure 3c-d**). The spin-coated device reached a modest EQE of 1.3% and a luminance of 2946 cd/m² at 8 V. The blade-coated devices show similar J-V-L curves as the spin-coated alternatives. Due to the monolayer coverage, the device with a QD film fabricated at 3 mm/s exhibits the lowest luminance and EQE: monolayer coverage provides less active material to capture charge carriers, leading to current leakage, decreased exciton generation, and photon emission. With increased surface coverage, blade-coated devices exhibit higher EQE, primarily due to suppression of current leakage and higher QD density, which enhances the likelihood of radiative recombination, especially at low current densities. In theory, devices with thicker QD film can absorb more electrons and holes, resulting in a higher number of excitons and higher luminance, but it is also limited by the carrier mobility and accessibility to the QDs. When the blade speed increases past 14 mm/s, the QD layer becomes too thick, leading to higher series resistance and drive voltage, which limits efficient QLED operation (**Figure S6**). All the blade-coated devices demonstrated good reproducibility, with a peak EQE observed at 7 mm/s (**Figure 4e**). The QD layer coated at this speed achieved the highest average EQE of 1.5% and an average luminance of 2350 cd/m² among the devices tested. Consequently, we selected this coating speed for further investigation.

All blade-coated CdSe-QLEDs were fabricated using the optimized blade speed for the QD layer (**Figure S7**). The J-V-L and EQE-L characteristics of a typical device are presented in **Figure 5a,b**. The devices exhibit high reproducibility, as demonstrated by the histogram of the peak EQEs of 20 devices (**Figure 5c**). The peak of the histogram indicates that most of the devices have an EQE of around 1.4%, comparable to that of the spin-coated alternatives. This consistent performance across devices highlights the reliability of the blade coating method in achieving uniform and optimally thick active layers, making it a viable alternative to traditional spin coating for scalable device fabrication. Additionally, only one-fourth of the active material solution is needed for each layer compared to that required for spin coating. This low-waste and cost-effective approach is promising for future mass production.

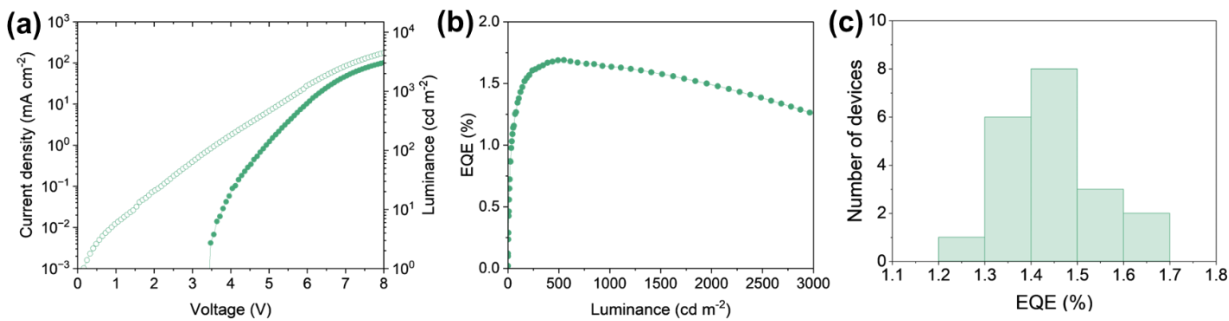

**Figure 5**. (a) The J-L-V curve and (b) EQE of the all-blade-coated devices. (c) Histograms of the peak EQEs for the all-blade-coated CdSe-QLED.

We sought to demonstrate the generality of this technique by applying it to lab-made InP/ZnSe/ZnS QDs that are engineered for performance in electrically driven devices. We synthesized red-emissive InP QDs with a thick ZnSe shell and thin ZnS shell to insulate the emissive core, decrease QD-QD electronic coupling and maximize the colloidal stability and processability of the QD inks, as detailed in the supplementary information (see **Figure S8** for additional solution phase QD photophysical behavior). Importantly, we demonstrate that this technique is amenable to other QD sizes and compositions while keeping other active layers the same. The size (~8 nm) and morphology of the QD core are uniform, which is essential for PL and EL (**Figure S9**). The shelled InP QDs reached a solution PLQY of 74%, much higher than the commercial CdSe/ZnS QDs. The morphologies of spin- and blade-coated QDs on ITO/PEDOT/TFB substrates are studied (**Figure S10**). All QD films have low roughness (Sq < 1.5 nm) and present a monolayer structure at 0.5 mm/s, with a second layer appearing at 1 mm/s. However, the layer-by-layer structure is indistinguishable as the coating speed increases. We note that the InP/ZnSe/ZnS QDs were engineered with an entropic ligand system,[41] consisting of ~82% oleate and ~18% 2-ethyl hexanoate species, unlike the single-type ligand in the commercial alternatives (**S1.3** and **Figure S11**). These ligands change the surface energy of the QD and competition between $W_{QD\text{-subs}}$ and $W_{QD\text{-}QD}$. Thus, the underlying physicochemical interactions require further investigation. Furthermore, we fabricated QLEDs with InP QD layers blade-coated at different speeds and fully spin-coated devices (**Figure S12**). The spin-coated InP-QLEDs reached an EQE as high as 7.4% and luminance higher than 10,000 cd/m$^2$, while the blade-coated devices at 7 mm/s exhibit an EQE of 6.6% and luminance over 10,000 cd/m$^2$. The devices show sharp EL spectra (full width at half maximum ~ 50 nm) that don't shift at different voltages. The results demonstrate the potential for the fabrication of high-performance blade-coated QLEDs, and highlight the need for improved understanding of interfacial interactions to achieve precise control of QD assembly for controlled thickness high-performance blade-coated QLEDs.

**Conclusion**

In conclusion, using fluid mechanics principles, we unraveled the origins of QD assembly from solution to enable the fabrication of QLEDs *via* a large-area deposition method, namely blade-coating. Coating speed critically impacts and determines QD active layer morphology and thickness, impacting device characteristics. By using commercial CdSe/ZnS QDs, we demonstrated that an optimal coating speed of 7 mm/s produces a QD film with approximately 163% surface coverage, a low roughness of 1.57 nm, and an enhanced EQE of 1.5%. Furthermore, all-blade-coated CdSe-QLEDs were fabricated by applying this optimized coating speed for the CdSe/ZnS QD layer with substantially less active material compared to spin-coated controls. This technique is amendable to lab-made InP/ZnSe/ZnS QDs with a much higher PLQY. High-performance blade-coated InP-QLEDs with an EQE approaching 7% and luminance over 10,000

cd/m$^2$ were demonstrated. Future work will focus on manipulating the surface chemistry of lab-made QDs to modify their interaction with the substrate, enabling better control over QD morphology and thickness. These findings provide valuable insights for developing scalable, cost-effective manufacturing processes for next-generation QLED technologies.

**Experimental Section**

**Materials.** Zn(Ac)$_2$·2H$_2$O (99.999% trace metals basis), Mg(Ac)$_2$·4H$_2$O (⩾98%), ethanolamine (99.0%), Ethanol (⩾99.5%), DMSO (⩾99.9%), octane (⩾99%), o-xylene (97%) and ethyl acetate (⩾99.5%) were purchased from Sigma-Aldrich. TMAH (98%) was purchased from Thomas Scientific. PEDOT:PSS (AI4083) was purchased from Heraeus Epurio. TFB (Mw ~100,000) was purchased from Montreal Optoelectronics Inc. CdSe/ZnS QD with octadecylamine ligands (PLQY~33%) was purchased from Ocean Nanotech. All materials were used as received without any further purification.

**Synthesis of ZnMgO NPs.** The ZnO NPs were prepared according to the method reported with some modifications.[42,43] TMAH (5 mmol) in ethanol (10 mL) was added dropwise to the flask containing Zn(Ac)$_2$·2H$_2$O (2.55 mmol) and Mg(Ac)$_2$·4H$_2$O (0.45 mmol) dissolved in DMSO (30 mL) under stirring at 700 rpm. The solution was stirred for 1 h in ambient conditions. ZnMgO NPs were precipitated by ethyl acetate and redispersed in ethanol. Ethanolamine (120 μl) was added to stabilize the NPs. The NPs were washed a second time and were redispersed in anhydrous ethanol for use.

**QD Morphology Sample Preparation.** The glass substrate (Ossila, quartz-coated glass) is cleaned with Hellmanex III water solution and isopropyl alcohol for 15 min each and then treated with plasma for 30 min. The ITO/PEDOT/TFB substrate is prepared by spin coating using the procedure in QLED device fabrication session. 10 μl of QD solution blade-coated on the treated glass and ITO/PEDOT/TFB substrate at different speeds.

**QLED Device Fabrication.** The ITO glasses (sheet resistance ~20 Ω/square) were cleaned by ultrasonic wash in Hellmanex III water solution and isopropyl alcohol for 15 min each and then treated with plasma for another 30 min. For the spin-coating process, 75 μl of PEDOT:PSS (filtered through a 0.45 μm PES filter) were spin-coated onto ITO glassed with multi-steps of 500 rpm (5 s) and 3,000 rpm (50 s), followed by annealing at 150 °C for 15 min. Then, the PEDOT:PSS-coated substrates were transferred into a nitrogen-filled glove box. 40 μl of TFB (8 mg/ml in o-xylene) was spin-coated at 2,000 rpm and annealed at 150 °C for 30 min. 35 μl of commercial or lab-made QD solution (15 mg/ml in octane) was spin-coated at 3,000 rpm, followed by annealing at 120 °C for 30 min. 40 μl of ZnMgO solution (30 mg/ml in ethanol) was spin-coated at 2,000 rpm and annealed at 90 °C for 15 min. For the blade-coating process, all the active material solutions and annealing temperature and time are the same. The distance between the blade and substrate is 0.2 mm, and the active material solution volume is 10 μl. PEDOT:PSS was bladed at 5 mm/s with substrate heating at 55 °C. Then, the substrates were transferred into the glove box. TFB was bladed at 3 mm/s, commercial or lab-made QD solution at different speeds, and ZnMgO solution at 7 mm/s without heating. Finally, the Ag electrode (~100 nm) was deposited by a thermal evaporator with a rate of 1 Å $s^{-1}$ at high vacuum (~ $3 \times 10^{-6}$ bar). All active layers of the electron-only device (ITO/ZnMgO/QD/ZnMgO/Ag) and hole-only device (ITO/PEDOT/TFB/QD/MoO$_3$/Ag) were fabricated according to the procedure described above and the MoO$_3$ (~10 nm) was deposited by a thermal evaporator at 0.5 Å $s^{-1}$ at high vacuum (~ $3 \times 10^{-6}$ bar).

**Characterization.** Atomic force microscopy (AFM) was performed on the QD film surface using a Bruker Bioscope Resolve atomic force microscope in air. The topographic images were collected under Tapping mode with Aluminum-coated tips (HQ:NSC19/Al BS, MikroMasch) with a spring constant of 0.5 N/m. The PL spectra were measured by Horiba Fluorolog Fluorometer. The J–V–L characteristics of the QLEDs were analyzed using a Keithley 2400 source meter to record the J-V curve and a Keithley 6500 multimeter with a calibrated Thorlabs Si switchable gain detector (PDA100A2) to record luminance under ambient conditions. The electroluminescence spectra were obtained with an Ocean Optics Flame spectrometer. The EQE was calculated by assuming that the emission obeys a Lambertian profile. The contact angles were measured by a ramé-hart 210 goniometer.

## ASSOCIATED CONTENT

**Supporting Information:** Supplementary figures showing the morphologies of ITO/PEDOT/TFB and CdSe/ZnS QD coated on ITO/PEDOT/TFB films, contact angles of QD solution on different substrates, a summary of AFM surface roughness, the estimated surface coverage at different coating speeds, J-V curves of electron- and hole-only devices, PL spectrum of QD film, EL spectra of CdSe-QLED, the morphology of QD film coated at 14 mm/s, CdSe-QLED performance with QD layer coated at 14 mm/s, photos of all-blade-coated CdSe-QLED, InP/ZnSe/ZnS synthesis and experimental details, solution phase QD photophysical behavior, QD physical characteristics, morphologies of InP/ZnSe/ZnS QD films coated at different speeds, $^1$H NMR spectrum of ligands on the InP/ZnSe/ZnS QD, schematic of InP-QLED structure and spin-coated and blade-coated device performance.

## AUTHOR INFORMATION

**Corresponding Author**
\* elr420@lehigh.edu

**ORCID**

*Yiman Xu: 0009-0005-1871-9181*
*Grant J. Dixon: 0009-0003-9663-455X*
*Qing Xie: 0000-0001-6996-4543*
*James F. Gilchrist: 0000-0003-2066-750x*
*Brandi M. Cossairt: 0000-0002-9891-3259*
*David S. Ginger: 0000-0002-9759-5447*
*Elsa Reichmanis: 0000-0002-8205-8016*

**Author Contributions**

Y.X., J.G., and E.R. conceived the project and its essential ideas. Y.X. set up the blade-coating system in the glove box, synthesized the ZnMgO nanoparticles, performed film processing and characterization, device fabrication and characterization, and prepared the manuscript. G.D. synthesized InP/ZnSe/ZnS QDs and performed characterizations. Q.X. provided technique support and the analysis tool for AFM. B.C. and D.G. engaged in intellectual discussions. All authors contributed to editing the manuscript.

**ACKNOWLEDGMENTS**

This work was supported by the National Science Foundation under the STC Grant No. DMR-2019444, for the Center for Integration of Modern Optoelectronic Materials on Demand (IMOD). In addition, E.R., Y.X., J.G. acknowledge partial support from Lehigh University, funds associated with the Carl Robert Anderson Chair in Chemical Engineering, and access to the Lehigh University Institute for Functional Materials and Devices Materials Characterization Facility (MCF) and Integrated Nanofabrication and Cleanroom Facility (INCF). The authors also appreciate assistance from Helen Larson for Transmission Electron Microscopy analysis and access to the Xiaoji Xu atomic force microscopy facilities.

# Supporting Information

# Landau-Levich Scaling for Optimization of Quantum Dot Layer Morphology and Thickness for Enhanced Quantum-Dot Light-Emitting Diode Performance

*Yiman Xu[1], Grant J. Dixon[2], Qing Xie[3], James F. Gilchrist[1], Brandi M. Cossairt[2], David S. Ginger[2], Elsa Reichmanis[1]\**

1. Department of Chemical and Biomolecular Engineering, Lehigh University, 124 E. Morton Street, Bethlehem, Pennsylvania 18015, United States

2. Department of Chemistry, University of Washington, Seattle, Washington 98195, United States

3. Department of Chemistry, Lehigh University, 6 E Packer Avenue, Bethlehem, Pennsylvania 18015, United States

*Email Address: elr420@lehigh.edu

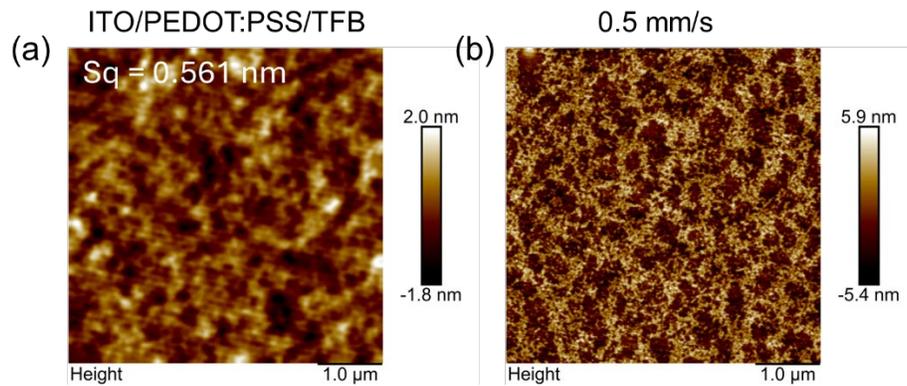

**Figure S1.** The morphologies of (a) ITO/PEDOT/TFB film, and (b) CdSe/ZnS QD coated on ITO/PEDOT/TFB at a speed of 0.5 mm/s.

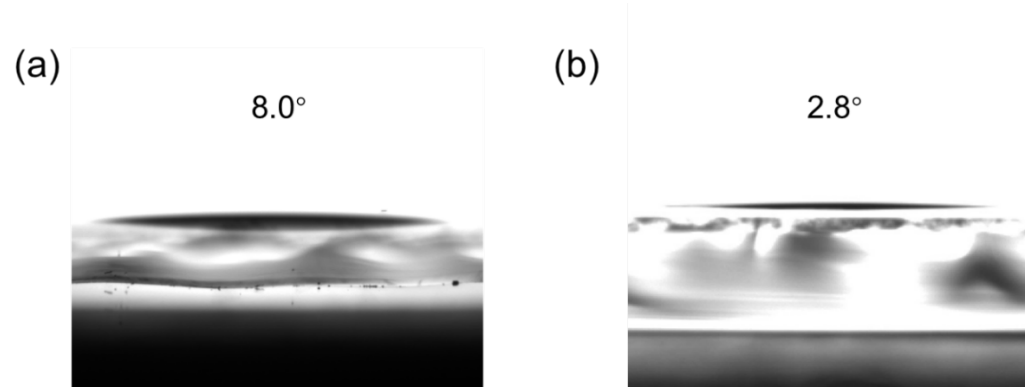

**Figure S2.** The contact angles of QD solution on (a) ITO/PEDOT/TFB and (b) glass substrate.

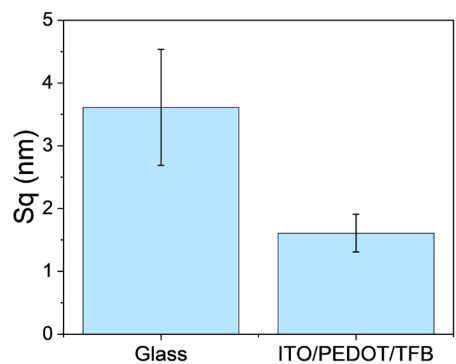

**Figure S3.** Summary of AFM surface roughness (Sq) for QDs on glass and ITO/PEDOT/TFB substrates.

**Table S1**. The estimated surface coverages at different coating velocities.

| Coating velocity u (mm/s) | Surface coverage c (%) | |
|---|---|---|
| 1 | 45.16 | |
| 2 | 71.69 | |
| 3 | 93.94 | 1st layer |
| 4 | 113.80 | |
| 5 | 132.05 | |
| 6 | 149.12 | |
| 7 | 165.25 | |
| 8 | 180.64 | |
| 9 | 195.40 | 2nd layer |
| 10 | 209.61 | |
| 11 | 223.37 | |
| 12 | 236.71 | |
| 13 | 249.68 | |
| 14 | 262.32 | |
| 15 | 274.67 | |
| 16 | 286.75 | |
| 17 | 298.57 | 3rd layer |
| 18 | 310.17 | |

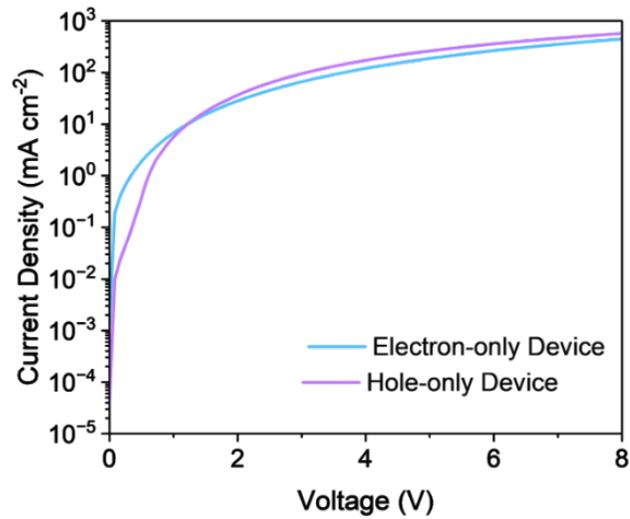

**Figure S4.** J-V curves of electron-only device (ITO/ZnMgO/QD/ZnMgO/Ag) and hole-only device (ITO/PEDOT/TFB/QD/MoO$_3$/Ag) using CdSe/ZnS QDs.

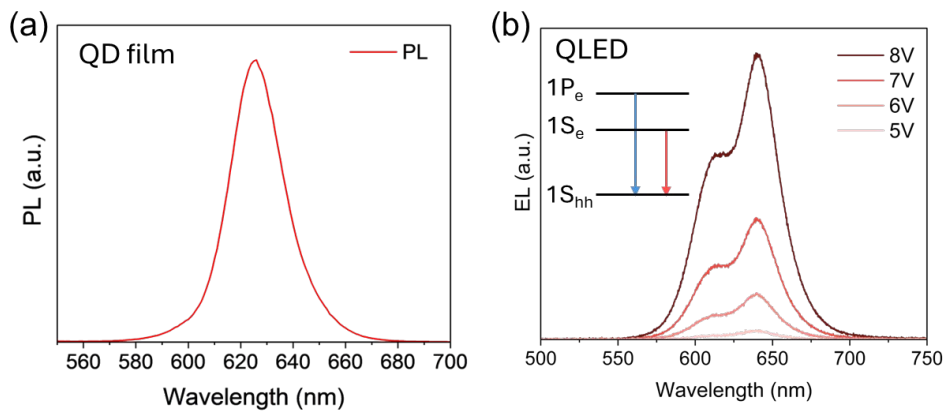

**Figure S5.** (a) PL spectrum of commercial CdSe/ZnS QD film. (b) EL spectra of CdSe-QLED device driven at different voltages. The insert explains the origin of the dual emission peaks.

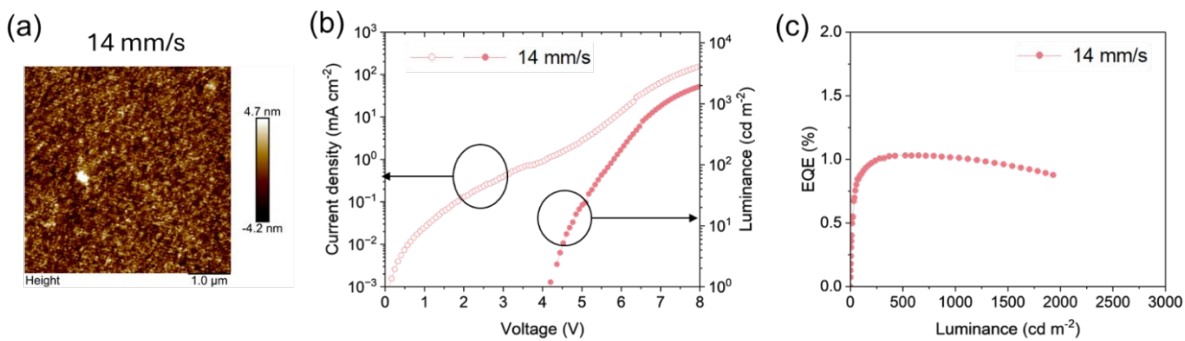

**Figure S6.** (a) Morphology of CdSe/ZnS QD film coated on ITO/PEDOT/TFB substrate at a speed of 14 mm/s. (b) J-V-L curves and (c) EQE-L curve for the CdSe-QLED device with a QD layer bladed at 14 mm/s.

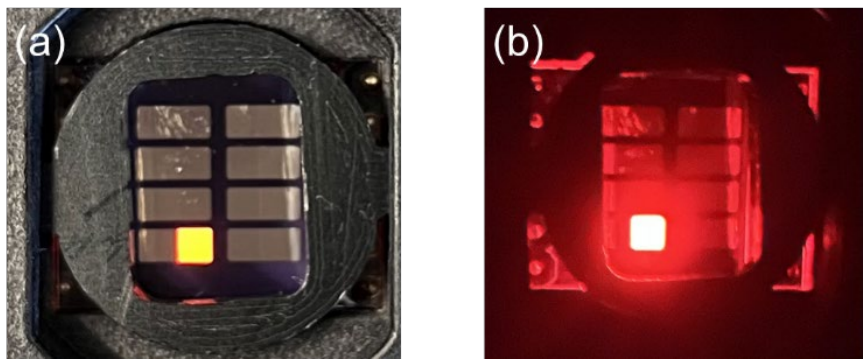

**Figure S7.** Photos of an all-blade-coated CdSe-QLED device (driven by 7V) under (a) ambient light and (b) in a dark box.

# InP/ZnSe/ZnS Synthesis and Experimental Details

## S1.1   Materials and Precursor Syntheses

### General Considerations

Unless otherwise noted, all reactions were carried out under an inert atmosphere of nitrogen using standard Schlenk line and glovebox techniques. All glassware was dried in a 160ºC oven overnight before use. Reactions were heated using J-KEM Gemini temperature controllers and hemispherical glass fiber heating mantles or a bath of silicon oil (<160ºC). Squalane and hexadecane were dried and degassed under dynamic vacuum at 100ºC for 16h before storing over 3Å sieves in a nitrogen-filled glovebox. All additional solvents used for synthesis and purification were purified using the method of Grubbs and stored in a glovebox.[1] Selenium, Sulfur, In(OAc)$_3$ and Zn(OA)$_2$ powders and solutions of TOP-Se and TOP-S were stored in a nitrogen-filled glovebox.

### Materials

Indium (III) Acetate (99.99%) was purchased from Strem and stored in a nitrogen glovebox. Zinc (II) oxide (99.99%), zinc (II) acetate (99.99%), tri-n-octyl phosphine (TOP) (97%), oleic acid (tech grade, 90%), myristic acid (≥99%), selenium powder (200 mesh, 99.99%), sulfur powder (99.98%), Hydrofluoric acid (HF) (ACS reagent, 48 wt%), hexadecane (99%), acetone (>99.9%), anhydrous ethyl acetate (99.8%), anhydrous acetonitrile (99.8%), anhydrous ethanol (>99.5%), and anhydrous octane (>99%) were purchased from Millipore Sigma and used as received. Zinc 2-ethyl hexanoate (20 wt% Zn), isopropanol (ACS reagent, 99.6%), methanol (99.9%), toluene (99.7%), and anhydrous diethyl ether (99.9%) were purchased from ThermoFisher Scientific. Oleic acid (>99%) and squalane (>98%) were purchased from TCI. Deuterated toluene was obtained from Cambridge Isotope Laboratories, dried over CaH$_2$, and distilled before use.

### Synthesis of Tris(trimethylsilyl)phosphine [P(TMS)$_3$]

*Caution! P(TMS)$_3$ is highly pyrophoric and generates toxic phosphine gas upon exposure to moisture. Handle and quench with care.*

Tris(trimethylsilyl)phosphine was prepared according to the literature, distilled, and stored in the dark inside a nitrogen-filled glovebox.[2]

**Synthesis of 0.4 M Indium Myristate Solution [0.4M In(MA)$_3$].** To prepare 0.4M In(MA)$_3$, myristic acid (16.99 g, ~19.7 mL, 74.4 mmol, 3.1 equiv.) and indium acetate (7.01g, 24.0 mmol, 1.0 equiv.) were combined and heated to 150ºC under nitrogen flow for 30 minutes. Then, the heat was reduced to 120ºC, and the flask placed under dynamic vacuum for 16h. Hexadecane (40.3 mL) was then added with gentle heating to dissolve the indium myristate, and the resulting waxy solid was stored in a nitrogen-filled glovebox. The solid can be gently warmed to 80-90ºC for dispensing.

**Synthesis of InP Single-Source Precursor Solution (InP SSP).** The InP precursor solution was prepared according to a modified procedure for TOP-Cluster$_{30}$.[3] All parameters were kept the same, but oleic acid (99%) was used in place of stearic acid, and the scale was increased to 4.5 mmol/Indium acetate. The resulting solution with an indium concentration of ~0.25 M was stored in a glovebox at -20ºC.

**0.15 M In(MA)$_3$/TOP.** A flask charged with 0.4M In(MA)$_3$ (11.25 mL, 4.5 mmol, *1.0 equiv.*) and hexadecane (6.75 mL) was evacuated and heated to 120ºC. The flask was then backfilled and heated to 150ºC, at which point TOP (12.0 mL, 6.0 equiv.) was introduced and allowed to react for 5 mins. The reaction was then cooled to room temperature, and the clear solution was transferred to a clean vial in a glovebox for storage.

**0.075 M P(TMS)$_3$.** In a glovebox, an oven-dried 50 mL Schlenk flask was charged with hexadecane (29.5 mL). Then, neat P(TMS)$_3$ (654 µL, 2.25 mmol) was measured with a microliter syringe and added to the hexadecane dropwise with stirring.

**Synthesis of InP Feeder Solution (InP Feeder).** The feeder solution was prepared fresh at the time of synthesis by combining equal volumes of the *0.15 M In(MA)$_3$/TOP* and *0.075 M P(TMS)$_3$* solutions in a nitrogen-filled glovebox.

**Synthesis of Zinc Oleate [Zn(OA)$_2$] powder and Preparation of 0.5 M Suspension.** Zinc oleate was synthesized using a modified version of the iron oleate from iron trifluoroacetate synthesis *via Hendricks et al.*[4] Lead (II) oxide is exchanged for Zinc (II) oxide (7.3 g, 90 mmol). The resulting powder was dried under dynamic vacuum for 24 hours to yield a free-flowing white powder in 93% yield (52.75g). $^1$H NMR (500 MHz, d$_8$-Tol) δ= 0.96 (t, $^3J_{H-H}$= 7 Hz, 6H, CH$_3$), 1.15-1.51(m, 40H, (CH$_2$)$_6$ and (CH$_2$)$_4$), 1.59 (m, 4H, COCH$_2$CH$_2$), 2.06-2.18 (m, 8H, =CHCH$_2$), 2.26 (t, $^3J_{H-H}$ =7.6 Hz, 4H, COCH$_2$), 5.50 (m, 4H, =CHCH$_2$).

*Preparation of 0.5 M Zn(OA)$_2$ suspension-* Zn(OA)$_2$ suspensions for shelling reactions were prepared by vigorously mixing the appropriate amount of Zn(OA)$_2$ powder in the corresponding amount of dry squalane for ~10 min (rt). Syringes used to add this suspension were also charged with the suspension under vigorous stirring to minimize settling of the powder.

**Preparation of 2.0 M HF in acetone.** The HF solutions for surface oxide treatment were prepared immediately before use by diluting 48 wt% HF (145 µL, 8 mmol) with acetone (1855 µL) in a 15 mL PTFE centrifuge tube. *Caution! HF is extremely toxic, corrosive, and should be handled with care.*

**Synthesis of 2.24M Tri-*n*-octylphosphine Selenide (TOP-Se) and 2.24M TOP-S Tri-*n*-octylphosphine Sulfide (TOP-S).** Saturated (2.24M) TOPSe and TOP-S were synthesized according to the literature[5] and stored in a nitrogen filled glovebox.

### S1.2  Synthesis of InP QD Cores

**Heat-Up Synthesis of Small InP QDs (Core 1)**

An oven-dried 50 mL 3-neck flask outfit with a stir bar, glass thermowell, reflux condenser, and rubber septum was cycled onto the Schlenk line 3x. Meanwhile, the InP SSP was thawed and a volume containing ~1.0 mmol/Indium (4.0 mL, ~1.8 g) was loaded into a syringe. The reaction flask was then charged with 0.4M In(MA)$_3$ solution (8.4 mL, 3.3 mmol, *1.0 equiv.*), evacuated, and heated to 120ºC. After backfilling the flask and heating to 150ºC, TOP (8.9 mL, 19.8 mmol, *6.0 equiv.*) was introduced. After 5 minutes, the flask was set to rapidly heat to 240ºC. The InP SSP-charged syringe was removed from the box and rapidly injected into the reaction upon the temperature reaching 190ºC. The reaction was maintained at 240ºC for 5 minutes, at which point the absorbance features stopped changing. This step affords core 1, which typically has a lowest-energy electronic transition between 510-530 nm (2.8-2.9 nm).

**Controlled growth of small InP QDs to Large InP QDs (Core 2)**

The small Core 1 can be grown to larger sizes by the slow addition of more indium and phosphorus precursors *via* the InP Feeder Solution. After the growth of core 1 appeared to cease, the syringe containing InP Feeder solution was retrieved from the glovebox. Starting 25 minutes after the initial InP SSP injection, the InP feeder solution was added at a slow dropwise rate (~1.0 mL min$^{-1}$) by hand in 1.0 mL increments every 10 minutes. Additions of InP Feeder were repeated in 10 min/ 1.0 mL increments until the desired size was reached.

**Purification of InP QD Core 2**

After reaching the desired size, the reaction was allowed to continue for 10 minutes at 240°C. Then, the reaction was removed from heat and allowed to cool in air. Volatiles were removed under reduced pressure (rt), and the flask was sealed and brought into the glovebox. To each 5.0 mL of the crude reaction were added anhydrous EtOAc (40 mL) and anhydrous MeCN (5 mL) to precipitate the QDs. The flocculated QDs were isolated by centrifugation (7830 rpm, 5 min), and the resultant dark red, oily pellets were resuspended in minimal anhydrous toluene. The QDs were then flocculated with excess anhydrous MeCN, centrifuged, and resuspended in toluene. Redissolution and precipitation with toluene/MeCN was repeated 3x to yield a dark red, waxy, solid pellet. The final pellets were combined and diluted to 10.0 mL in anhydrous toluene to generate a stock solution for storage.

**S1.3    Synthesis of InP/ZnSe/ZnS Core/Shell QDs**

**General Overview and Scheme**

The shelling methodology presented herein is a close adaptation of the methods laid forth by Choi *et al.*[6] The general approach towards heteroepitaxial shelling of the ZnSe and ZnS layers on InP consists of 1) An exchange of surface indium carboxylates with zinc carboxylates, followed by etching of residual surface oxides with HF; 2) saturation of ZnSe species on the surface at low temperature, followed by the high-temperature transformation of ZnSe to a crystalline shell; 3) low-temperature introduction of S precursor to saturate the InP/ZnSe surface, followed by the same high-temperature annealing process. Reaction stoichiometry was determined by consideration of the core size and desired shell thickness (See **S2.1**).

**Removal of InP Core 2 Surface Oxides with HF and Indium to Zinc Exchange**

A 100 mL oven-dried 3-neck flask was cycled onto the Schlenk line 3x and left under vacuum for 30 mins. In the meantime, Zn(OA)$_2$ suspensions and 4.0 M HF in acetone were prepared (see **S1.1**). The flask was backfilled and charged with anhydrous squalane (20 mL). Then, an amount of the InP QD Core 2 stock solution containing 440 nmol of QDs (1575 μL, see **S2.1**) was drawn into a syringe, injected into the flask and the flask was evacuated as it was heated to 110°C for 30 min. The flask was then backfilled, the 0.5 M Zn(OA)$_2$ suspension (3.0 mL, 1.5 mmol) added *via* syringe, and the flask was then heated to 200°C under nitrogen flow. After 20 mins at 200°C, the flask was cooled to 120°C. The flask stopcock was closed to seal off the flask from the Schlenk line, and the rubber tubing connecting the flask to the line was placed under dynamic vacuum. Then, 2.0 M HF in acetone (0.4 mL, 0.8 mmol) was rapidly injected at 120°C, and the stopcock was immediately (and carefully) cracked open to evacuate the reaction headspace. After 2 minutes, the atmosphere was backfilled with nitrogen.

**Growth of Thick ZnSe shells on InP Core 2**

Once the nitrogen atmosphere was restored after introducing HF, an amount of TOPSe (715 μL, 1.6 mmol) targeting a 3.5 nm thick ZnSe shell was introduced to the reaction at 120ºC. The temperature was then ramped to 260ºC and held for 10 minutes. Then, the reaction was heated to 335ºC. Once at 335ºC, additional $Zn(OA)_2$ suspension (10.0 mL, 5.0 mmol) was added dropwise over 10 minutes to avoid large temperature fluctuations. After completing the $Zn(OA)_2$ addition, the flask was held at 335ºC for 30 minutes.

**Growth of Thin ZnS Shells on InP/ZnSe**

The reaction containing InP/ZnSe QDs at 335ºC was removed from heat after 30 minutes and cooled to 120ºC. Then, an amount of TOPS (161 uL, 0.36 mmol) needed to grow a 0.3 nm thick shell on InP/ZnSe was introduced, and the reaction heated to 260º. After 10 minutes at 260ºC, the flask was heated to 335ºC and held for another 30 minutes. The flask was then removed from heat, cooled with compressed air, evacuated to remove volatiles, and brought into the glovebox for purification.

**Purification of InP/ZnSe/ZnS**

The crude reaction solution (~20 mL) containing InP/ZnSe/ZnS QDs was diluted with anhydrous toluene (5.0 mL) and centrifuged (7830 rpm, 5 min) to remove insoluble material. Then, each 10.0 mL portion of the supernatant was precipitated by adding 40 mL of anhydrous EtOH and centrifuged (7830 rpm, 5 min). The resulting hazy red supernatant was discarded, and the bright red, oily pellet was resuspended in minimal toluene. The QDs were precipitated once more with EtOH to yield a bright red, dry, powdery pellet that was readily soluble in nonpolar solvents.

**Ligand Exchange to Entropic, Mixed Short/Long Chain Ligand Sphere.**

To improve both the colloidal processability[7] and charge injection properties[8] of the QDs, an entropic ligand sphere of short and long-chain ligands was constructed by first suspending the QD solution in anhydrous hexadecane (15.0 mL) and heating to 120ºC under reduced pressure. In a separate flask, zinc 2-ethyl hexanoate (131.5 μL, 0.4 mmol), anhydrous hexadecane (5.0 mL) and TOP (180 μL, 0.4 mmol) were mixed and heated to 100ºC for 5 minutes. The zinc solution was transferred to the QDs by syringe, and the reaction heated to 200ºC for 30 minutes. The flask was then cooled to room temperature, and zinc acetate (73.4 mg) was added under nitrogen flow. The flask was then evacuated and heated to 120ºC for one hour. The chalky solution containing mostly insoluble material was then cooled and brought into the box, and QDs precipitated by adding EtOH (40 mL) to each 10 mL of the reaction with vigorous mixing. The pellets collected from centrifugation were then transferred to a clean flask containing anhydrous hexadecane (20 mL). Zinc oleate powder (251.32 mg, 0.4 mmol) was transferred to the flask under nitrogen flow, which was then sealed and heated to 200ºC. After 30 minutes, the chalky solution had turned transparent again, and the flask was cooled to rt. The crude solution was worked up identically to the previous exchange, suspended in anhydrous pentanes, and passed through a 0.2 μM PTFE syringe filter into a pre-weighed glass vial. The pentanes were removed under reduced pressure overnight (16h) before measuring the isolated mass. After weighing the isolated QD powder, the QDs were suspended in anhydrous octane at a concentration of 15 mg/mL for device fabrication. We estimate a final ligand coverage of ~82% oleate and ~18% 2-ethyl hexanoate by $^1$H NMR analysis (**Figure S11**).

**S2.    QD Characterization Methods**

## S2.1 Sizing, Concentration Determination, and Stoichiometry Considerations

Concerning the optical sizing of pristine InP, the literature shows considerable inconsistencies regarding the calculated crystallite size at a given wavelength. Additionally, some of the most frequently employed sizing curves are incorrectly applied or based on incorrect sizing functions– a detailed discussion of which can be found in the supplementary material recently published by Calvin *et al.*[9]

Considering the lack of consensus regarding the optical sizing of InP nanocrystals, we made use of the sizing curve presented by Ministro.[10] Specifically, this sizing curve includes information on both size and extinction coefficient, allowing for self-consistent QD stock solution concentration and crystallite size for shelling stoichiometry calculations.

Stock solutions of QDs were analyzed as follows. After purification, InP QD solutions were dried to a powder under reduced pressure for 16 hours. If the sample remained oily, it was subjected to further purification to yield a powder before repeating the drying process. Then, in a glovebox, the QD powder was suspended in toluene (10.0 mL), stirred until dissolved, and then passed through a 0.45 µM PTFE syringe filter. A dilution was prepared to measure absorbance by adding 15 µL of QD stock to 4985 µL of anhydrous toluene in a glovebox, using gas-tight microliter syringes. This dilution affords absorbance traces with OD<2.0 to ensure optical linearity across the region of interest. This was repeated twice more for a total of three absorbance measurements. Then, the Ministro sizing curve allows for an average size determination from the three measurements after conversion of the spectra to eV.[11] The size is determined by

$$E_g = +\frac{1}{(0.119 \pm 0.003) \times D^2}$$

where $E_g$ is the energy of the lowest energy electronic transition (eV), and D is the diameter of the QDs (nm). The average of the 3 sizes obtained for one QD stock solution was considered for shelling stoichiometry.

The amount of QD stock solution containing 440 nmol of InP QDs ($V_{440\ nmol}$) can then be determined by applying the provided extinction coefficients at 335 nm and 410 nm

$$\varepsilon_{335} = (4.4 \times 10^4) \times D^3 \qquad \varepsilon_{410} = (1.29 \times 10^4) \times D^3$$

where $\varepsilon_{335}$ and $\varepsilon_{410}$ (L x M$^{-1}$ x cm$^{-1}$) are the extinction coefficients at 335 nm and 410 nm, respectively, and D is the QD diameter (nm). The average optical density of the 3 absorbance samples at each respective wavelength can then be used to back out the stock solution concentration, which is averaged between the results at each wavelength ($C_{avg}$). See **S2.2** for calculated values specific to the sample.

*Shelling Stoichiometry Calculations*- For shelling reactions, the concentration of zinc carboxylate was kept in large excess to maintain colloidal stability and minimize the effects of thermal decomposition at elevated temperatures. The chalcogenide precursors were treated as limiting reagents, and were determined as follows: Using the lattice parameter of zincblende ZnSe (0.556 nm) and zincblende ZnS (0.539 nm), the total amount of precursors for a batch of QDs (440 nmol) was determined by dividing the total desired shell volume for the batch of QDs by the atomic density of the shelling material.

## S2.2 Sample Calculations

| InP Core Property | Value |
|---|---|
| $\lambda_{LEET}$ | 577 nm *(2.15 eV)* |
| $OD_{avg,\ 335\ nm}$ | 1.27 |
| $\varepsilon_{335}$ | $1.5 \times 10^6\ L \times M^{-1} \times cm^{-1}$ |
| $C_{335}$ | $8.46 \times 10^{-7}\ mol \times L^{-1}$ |
| $OD_{avg,\ 410\ nm}$ | 0.37 |
| $\varepsilon_{410}$ | $4.4 \times 10^5\ L \times M^{-1} \times cm^{-1}$ |
| $C_{410}$ | $8.31 \times 10^{-7}\ mol \times L^{-1}$ |
| $C_{avg}$ | $8.39 \times 10^{-7}\ mol \times L^{-1}$ |
| $V_{440\ nmol}$ | 1575 μL |

## S2.3 Structural Characterization

Powder x-ray diffractograms were acquired on a Bruker D8 powder x-ray diffractometer equipped with a Pilatus 100K large area 2D detector and 1.0 mm collimator, using a Cu anode microfocus x-ray source (1.541Å). Samples for pXRD were prepared in a glovebox by drop-casting concentrated QD solutions onto clean Si substrates, dried in a vacuum overnight, and the resultant powdery films measured within 1h of air exposure. *All imaging* was done on a FEI Tecnai G2 F20 SuperTwin microscope operated at 200 kV using bright field imaging. Samples were prepared in a nitrogen glovebox by drop-casting 5 mL of dilute QD suspensions in toluene onto a suspended ultrathin carbon film on a lacey carbon support film, 400 mesh, copper grids purchased from Ted Pella Inc., allowed to dry fully (10 min), then placed under vacuum overnight. Samples were measured within 1 hour of air exposure. $^1$H-NMR spectra were recorded on a Bruker AVANCE NEO spectrometer (500 MHz) at 25 ºC using standard pulse techniques and a 30 second relaxation delay to allow for full relaxation and quantitative integrations of the QD sample. $^1$H-NMR samples were prepared at a concentration of ~30mg/mL in $d_8$–toluene after drying QD powders under vacuum for 16 hours. $^1$H-NMR data is reported as follows: chemical shift (ppm), multiplicity (s = singlet, t = triplet, m = multiplet), coupling constant (Hz), integration, and assignment.

## S2.4 Optical Characterization

*UV-Vis Absorbance-* Measurements were collected using either an Agilent Cary 60 or Agilent Cary 5000 Spectrophotometer in dual beam mode. Solution concentrations were adjusted to maintain OD<2 within regions of interest. All solution phase absorbance was collected in 1.0 cm fluorescence cuvettes (Spectrocell). Reactions were monitored by diluting 20 μL aliquots in 3.0 mL of toluene, followed by immediate measurement.

*Photoluminescence-* Standard emission spectra collected to monitor syntheses were obtained from 480-800 nm using a Horiba Jobin Yvon FluoroMax-4 Fluorescence spectrophotometer with an excitation wavelength of 465 nm, both emission and excitation bandwidths set to 1.0 nm, and integrated for 0.1 s with 1.0 nm resolution.

*Time Resolved Photoluminescence (TRPL) and Photoluminescent Quantum Yield (PLQY)-* TRPL and PLQY were measured on an Edinburgh FLS1000 Fluorometer equipped with a visible photomultiplier tube detector. All samples and references were prepared in 1.0 cm quartz fluorescence cuvettes using anhydrous toluene, sealed inside a glovebox atmosphere, and measured the same day. TRPL measurements were excited with an Edinburgh EPL-405 pulsed laser diode (excitation wavelength of 402.4 nm) operating with a pulse period of 2 µs. Signal was accumulated for 300s over 4096 channels across a 1 µs time range. Absolute PLQY measurements were collected in an Edinburgh FLS1000 integrating sphere (N-M01) and excited using a Xenon flash lamp at 465 nm with a slit width of 2.0 nm and referenced to solvent. All sample concentrations were adjusted to have OD<0.1 at 465 nm. Emission was collected with a 0.2 nm slit width from 455-720 nm and integrated for 0.5s with 0.1 nm resolution. PLQY integrations were done using the associated Fluoracle® software.

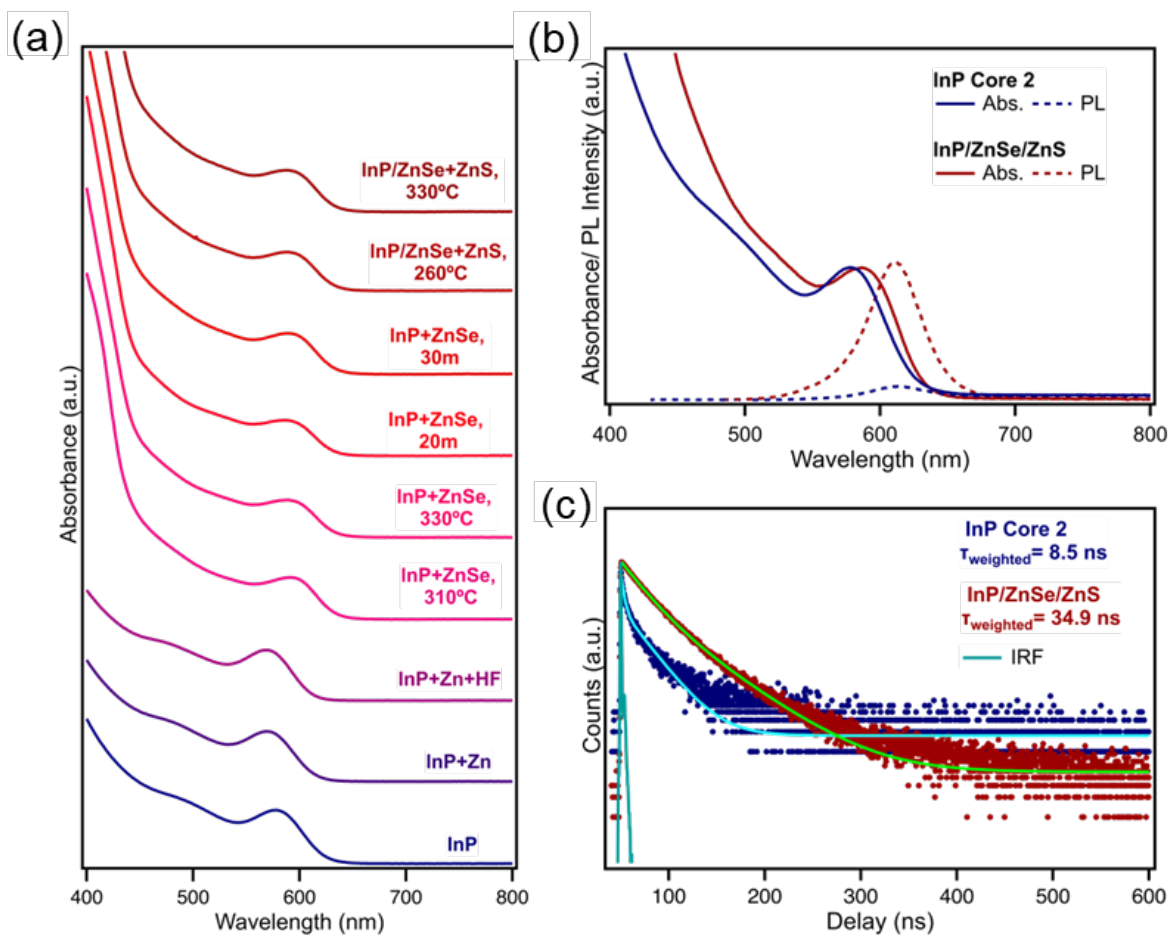

**Figure S8. Solution Phase InP/ZnSe/ZnS QD Photophysical Behavior** (a) Temporal evolution of the InP/ZnSe/ZnS absorbance spectra over the course of the shelling procedure; (b) Absorbance and photoluminescence of the InP Core QDs and InP/ZnSe/ZnS core/shell QDs; (c) Time-resolved photoluminescence decay of InP Core 2 and InP/ZnSe/ZnS. The decay profiles were fit to biexponential functions (light blue and green traces) to extract the weighted average lifetimes, reported in nanoseconds. Instrument response function included in teal.

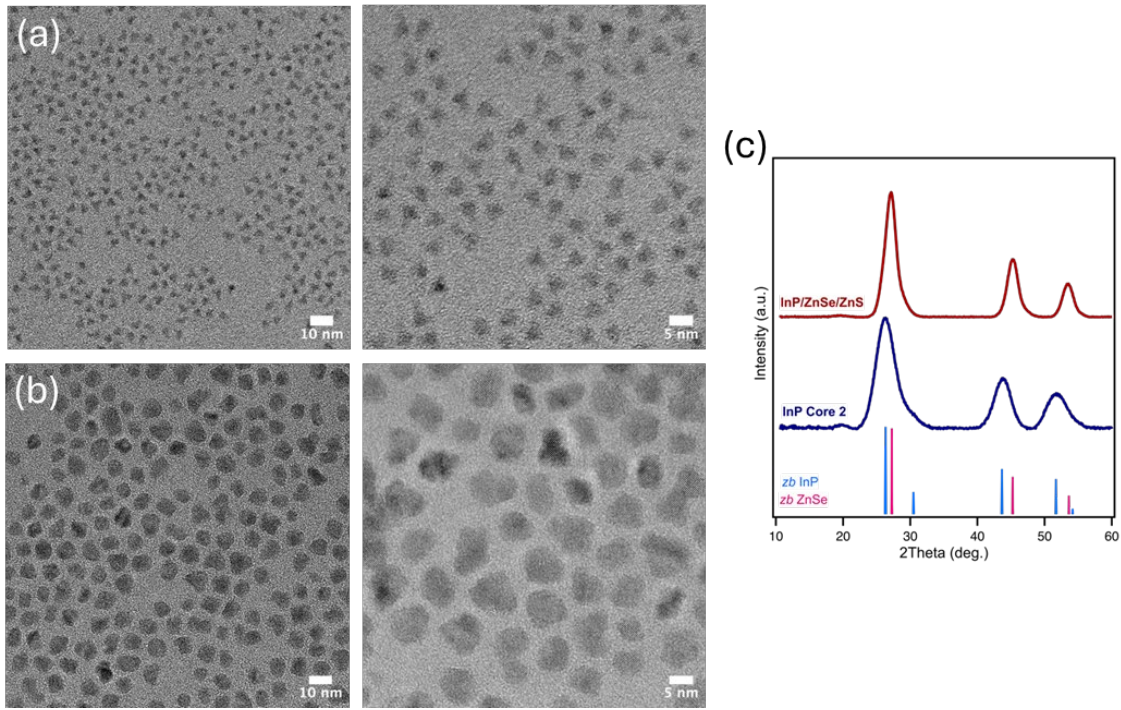

**Figure S9. InP/ZnSe/ZnS QD Physical Characteristics** (a) TEM micrographs of InP core 2 (~3.2 nm); (b) TEM micrographs of InP/ZnSe/ZnS core/shell QDs; (c) Powder XRD diffraction of InP Core 2 and InP/ZnSe/ZnS QDs. Reference patterns for zincblende InP and zincblende ZnSe included for reference.

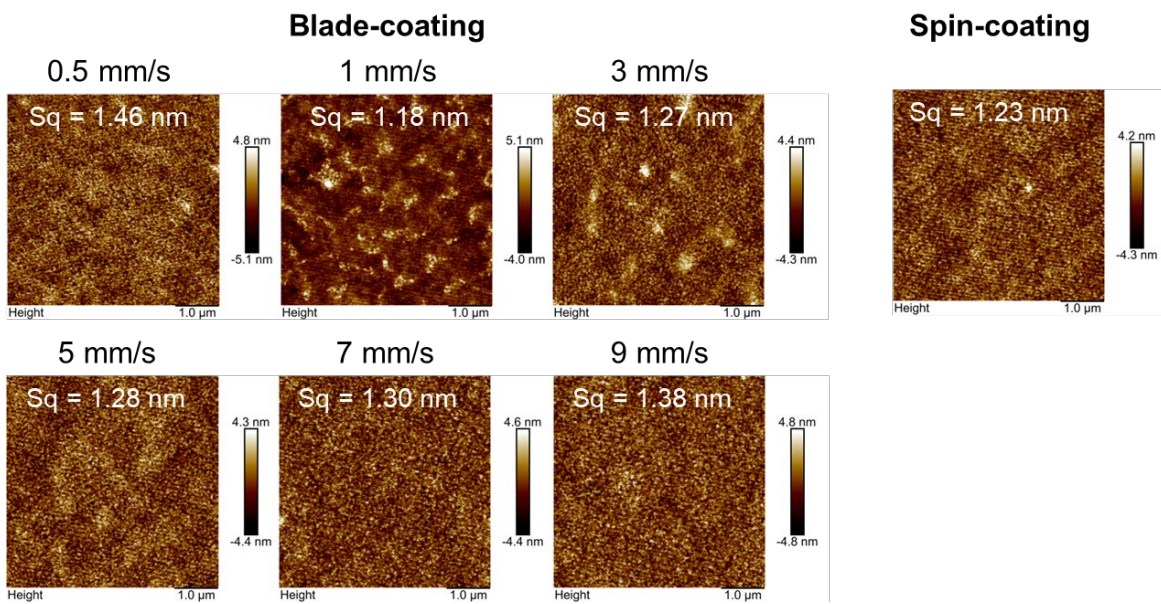

**Figure S10.** The morphologies of InP/ZnSe/ZnS QD films blade-coated at different speeds on ITO/PEDOT/TFB substrate with different roughness (Sq).

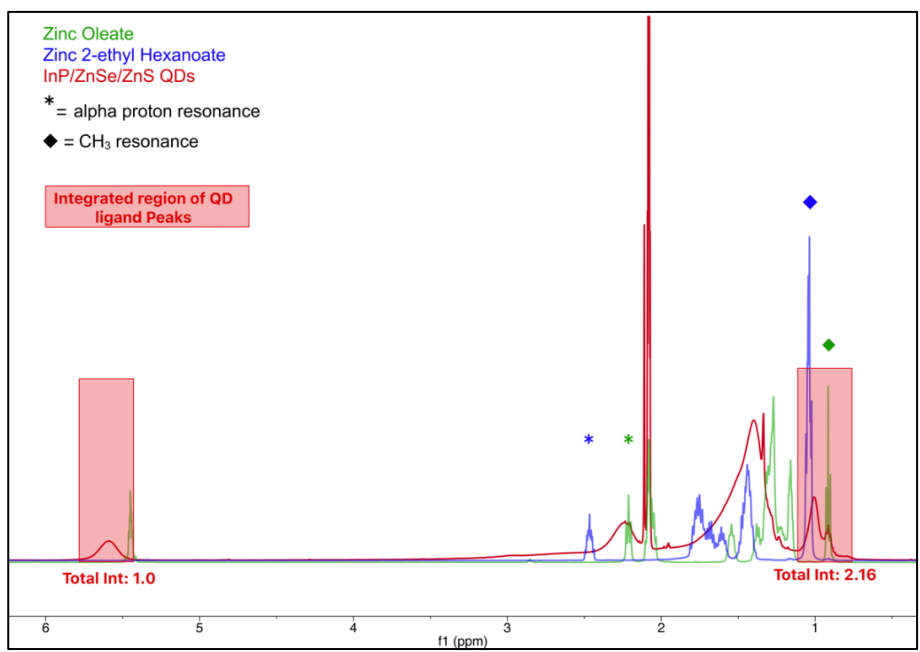

**Figure S11.** $^1$H NMR spectrum of oleate and 2-ethyl hexanoate capped InP/ZnSe/ZnS QDs (red) compared to zinc oleate (green) and zinc 2-ethyl hexanoate (blue). By normalizing the total integrated intensity of the methyl peaks (0.75~1.10 ppm) to the vinyl resonance of oleate (~5.6 ppm), we find an excess integrated intensity of 2.16-1.5 = (0.66 / 6 H's)*2 = 0.22 2-ethyl hexanoate ligands for every one oleate ligand, or an estimated oleate percentage of ~82%.

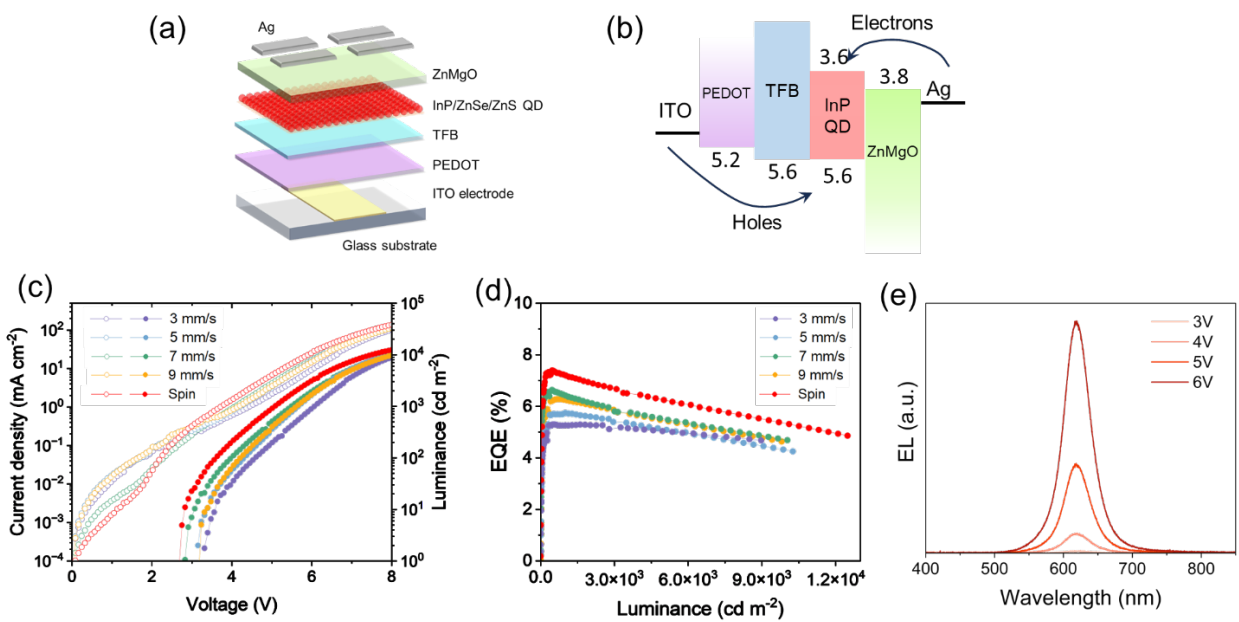

**Figure S12.** (a) The schematic of the InP-QLED structure and (b) the energy level diagram for active layers. (c) J-L-V and (d) EQE-L curve for QD layers spin-coated and blade-coated at different speeds. (e) The EL spectra of the InP-QLED driven at different voltages.